\documentclass[aps,prb,twocolumn,longbibliography,superscriptaddress,floatfix]{revtex4-2}

\usepackage{amsmath,amssymb,bm}
\usepackage{graphicx}
\usepackage{xcolor}
\usepackage{CJK}
\usepackage{comment}
\usepackage[colorlinks=true,linkcolor=blue,citecolor=blue,urlcolor=blue]{hyperref}
\usepackage[capitalize,nameinlink]{cleveref}

\newcommand{\avg}[1]{\langle #1 \rangle}
\newcommand{\Tr}{\mathrm{Tr}}

\newcommand{\eff}{\mathrm{eff}}
\newcommand{\ii}{\mathrm{i}}

\begin{document}

\title{Finite-temperature crossover from coherent magnons to energy superdiffusion in the PXP model}

\begin{CJK*}{UTF8}{}
\author{Shengtao Jiang (\CJKfamily{gbsn}蒋晟韬)}
\address{Stanford Institute for Materials and Energy Sciences, SLAC National Accelerator Laboratory and Stanford University, Menlo Park, CA 94025, USA}
\author{Jean-Yves Desaules}
\affiliation{Institute of Science and Technology Austria (ISTA), Am Campus 1, 3400 Klosterneuburg, Austria}
\author{Marko Ljubotina}
\affiliation{Physics Department, Technical University of Munich, TUM School of Natural Sciences, Lichtenbergstr. 4, Garching 85748, Germany}
\affiliation{Munich Center for Quantum Science and Technology (MCQST), Schellingstr. 4, M\"unchen 80799, Germany}
\author{Thomas Scaffidi}
\address{Department of Physics and Astronomy, University of California, Irvine, Irvine, California 92697, USA}

\date{\today}

\begin{abstract}
The PXP chain was recently shown to exhibit superdiffusive energy transport with Kardar-Parisi-Zhang-like scaling, $z\approx3/2$, joining a growing number of spin chains with this exponent. An understanding of how this anomalous hydrodynamics emerges from microscopics is, however, still lacking. In this work, we show that finite-temperature energy transport in this model provides a window into the emergence of superdiffusion. At finite temperature, the energy autocorrelation function exhibits a crossover from short-time coherent dynamics to long-time hydrodynamics. The short-time behavior is dominated by a single magnon band and can be understood analytically. In momentum space, this regime is characterized by spectral weight near $q=\pi$. The damping time $\tau$, which separates the short-time magnon-dominated behavior from the late-time hydrodynamics, grows rapidly upon cooling, consistent with an activated form $\tau(\beta)\sim \beta e^{\Delta\beta}$ with a gap scale set by the magnon band. At longer times, the spectral weight transfers to $q=0$ and the running decay exponent drifts toward the superdiffusive value $z=3/2$. Finite-temperature energy transport therefore provides a bridge between microscopic magnon physics and late-time superdiffusion in the PXP model.
\end{abstract}

\maketitle
\end{CJK*}

\section{Introduction}

Transport in low-dimensional quantum systems can exhibit a remarkably broad range of behaviors. In generic chaotic systems, conserved densities such as spin, charge, and energy are expected to relax diffusively at late times~\cite{Prosen_2012,Karrasch_2014,Lux2014,Friedman_2020,Bertini2021RMP}. Faster-than-diffusive transport is possible, but it typically requires special structure. Integrable systems, for example, can support ballistic transport because their extensive set of conserved quantities can prevent currents from decaying~\cite{Sutherland2004,Bertini2021RMP}. When additional symmetries are present, these systems may also exhibit transport that is intermediate between diffusion and ballistic propagation. The paradigmatic example is superdiffusive Kardar-Parisi-Zhang-like (KPZ) scaling, with dynamical exponent $z=3/2$, which has now been established in a growing number of spin-chain models, both theoretically and experimentally~\cite{Znidaric_2011,Spohn_2014,Ilievski_2018,Ljubotina_2017,Ljubotina_2019,De_Nardis_2019,De_Nardis_2020,DeNardisSolitons2020,Bulchandani_2020,Bulchandani_2020_2,Bulchandani_2021,Scheie2021,Wei2022KPZ,Dupont2021Crossover,Weiner2020KPZ,Rosenberg2024Science}. Anomalous hydrodynamic regimes have also been discussed beyond one-dimensional spin chains, notably in ultraclean electronic fluids~\cite{Lucas_2018,FritzScaffidi,Hui_2025}, which in two dimensions show superdiffusive current dynamics with $z=4/3$~\cite{ThuillierScaffidi2026AC}. This naturally raises the question: can robust faster-than-diffusive transport arise in nonintegrable models, and if so, what microscopic structures underlie it?


The PXP model occupies a special position in this discussion. It arises as an effective Hamiltonian for Rydberg atom arrays in the strong blockade regime and is also related to earlier work on hard-core bosons and even to lattice gauge theories~\cite{FendleySachdev,Lesanovsky2012,Labuhn2016,Bernien2017,Bluvstein2021,Surace2020}. Since the observation of long-lived revivals in Rydberg experiments~\cite{Bernien2017}, it has become the canonical setting for quantum many-body scars, characterized by a few specific initial states that avoid thermalization in an otherwise chaotic system~\cite{ChandranReview,Serbyn2021,MoudgalyaReview}.  
Recently, Ref.~\onlinecite{Ljubotina2023Superdiff} numerically demonstrated that the PXP chain exhibits superdiffusive energy transport at infinite temperature, with an effective dynamical exponent close to the KPZ value $z=3/2$. While it has since then been linked to the lattice gauge theory interpretation of the model~\cite{bhakuni2025LGT}, the microscopic mechanism behind this superdiffusion is still poorly understood. At the same time, a rather different line of work has emphasized coherent magnon dynamics as an organizing principle for understanding the quantum many-body scars of the same model~\cite{IadecolaMagnons}. These two perspectives---magnon physics and high-temperature anomalous hydrodynamics---have largely developed in parallel. A central motivation of the present work is to connect them with the broader goal of providing a microscopic understanding of how late-time superdiffusion can emerge from interacting magnons.

Finite temperature is a natural setting for doing so because it introduces a tunable crossover scale between microscopic and hydrodynamic behavior. As we will show, for times shorter than the thermalization time $\tau(\beta)$, the energy dynamics is controlled by the single magnon band, and the energy density autocorrelation function $C(x=0,t)$ follows a damped oscillatory $e^{-i \Delta t} / \sqrt{t}$ law, a signature of the quadratic magnon band minimum at momentum $\pi$.
At late times, the correlator becomes real and decays as a power law $t^{-1/z}$, with the running exponent approaching the superdiffusive KPZ value $z=3/2$ already observed at infinite temperature. Further, in momentum space, the spectral weight moves from $q=\pi$ to $q=0$. 
In this sense, finite temperature provides a unique window into how the system transitions from a microscopic coherent sector to long-wavelength superdiffusive transport.

This framing is closely related in spirit to the finite-temperature crossover perspective that has proved fruitful in the isotropic Heisenberg chain, where low-temperature and high-temperature dynamical regimes were brought into a unified picture through a temperature-dependent spatiotemporal crossover~\cite{DeNardisSolitons2020,Dupont2021Crossover,Weiner2020KPZ}. 
However, there are two important differences with the Heisenberg case: (1) the PXP chain is gapped and the crossover timescale $\tau(\beta)$ thus shows an activated behavior with temperature $\tau(\beta)\sim \beta e^{\Delta\beta}$ (with $\beta$ the inverse temperature and $\Delta$ the gap), as opposed to $\tau \propto \beta$ in the gapless Heisenberg chain; (2) the low-temperature coherent transport for the Heisenberg chain has a dominant ballistic ($z=1$) component~\cite{Dupont2021Crossover,steve-xxz}, as opposed to the $e^{-i \Delta t} / \sqrt{t}$ law we observed in the PXP chain.

This finite-temperature perspective also allows us to revisit the chemical-potential, or PNP, deformation of the PXP model. At infinite temperature, Ref.~\onlinecite{Ljubotina2023Superdiff} found that a positive deformation improves the convergence of the effective transport exponent toward the KPZ value. We show that this trend persists at finite temperature: a moderate positive deformation accelerates the approach of the running exponent toward $z=3/2$. 



\section{Model and Methods}
\label{sec:model}
We consider a chain of $L$ two-level atoms, where the local Hilbert space at site $x$ is spanned by $|g_x\rangle$ and $|r_x\rangle$, denoting the atomic ground state and Rydberg excited state, respectively. The Rydberg blockade restricts us to the constrained subspace in which no basis state contains adjacent excitations, i.e., no neighboring pair $|r_x r_{x+1}\rangle$.
We consider the PXP Hamiltonian
\begin{equation}
H_0 = \sum_x P_{x-1} X_x P_{x+1},
\end{equation}
where $X_x = |g_x\rangle\!\langle r_x| + |r_x\rangle\!\langle g_x|$ flips the two-level atom on site $x$ and $P_x = |g_x\rangle\!\langle g_x|$ projects onto the atomic ground state. The surrounding projectors implement the blockade constraint by allowing a flip on site $x$ only when both neighboring sites are in their ground state. Any state that initially has no neighboring excitations will not develop them over time. Throughout this work, we thus restrict the Hilbert space to the sector without such excitations. 
We also study the chemical-potential deformation
\begin{equation}\label{PNP}
H_\mu = H_0 + \mu \sum_x P_{x-1} n_x P_{x+1},
\end{equation}
where $n_j = |r_j\rangle\!\langle r_j|$ projects onto the Rydberg state.

As energy is the only conserved charge~\cite{park2025graph}, we study the transport of this quantity as previously done in Ref.~\onlinecite{Ljubotina2023Superdiff}. Our central probe is the connected finite-temperature energy density correlation function
\begin{equation}
\label{eq:correlator}
\begin{split}
C(x,t) &= \avg{h_x(t) h_0(0)}_\beta^c \\
&\equiv \avg{h_x(t) h_0(0)}_\beta - \avg{h_x(t)}_\beta \avg{h_0(0)}_\beta
\end{split}
\end{equation}
where $h_x=P_{x-1} \left(X_x+\mu n_x\right) P_{x+1}$ is the local energy density operator. Here $\langle\cdot\rangle_\beta$ denotes a trace with the thermal density matrix $e^{-\beta H}$:
\begin{equation}
\avg{\mathcal{O}}_\beta \equiv \frac{\Tr\!\left(\mathcal{P}e^{-\beta H}\, \mathcal{O}\mathcal{P}\right)}{\Tr\!\left(\mathcal{P}e^{-\beta H}\mathcal{P}\right)},
\end{equation}
where $\mathcal{P}$ projects out the states that do not satisfy the blockade constraint.
The effective time-decay exponent can be extracted from the autocorrelation function as
\begin{equation}
\frac{1}{z_\eff(t)} = -\frac{d\ln |C(x=0,t)|}{d\ln t}.
\end{equation}
For diffusive dynamics one expects $z=2$, while KPZ superdiffusion corresponds to $z=3/2$.
Lastly, we will also consider the momentum-space correlation function,
\begin{equation}
C(q,t)=\sum_x e^{-\ii q x} C(x,t).
\end{equation}
which will be useful to track the transfer of spectral weight from $q=\pi$ to $q=0$ as the coherent magnon dynamics gives way to hydrodynamics.

\begin{figure*}[t]
  \centering
  \includegraphics[width=0.7\textwidth]{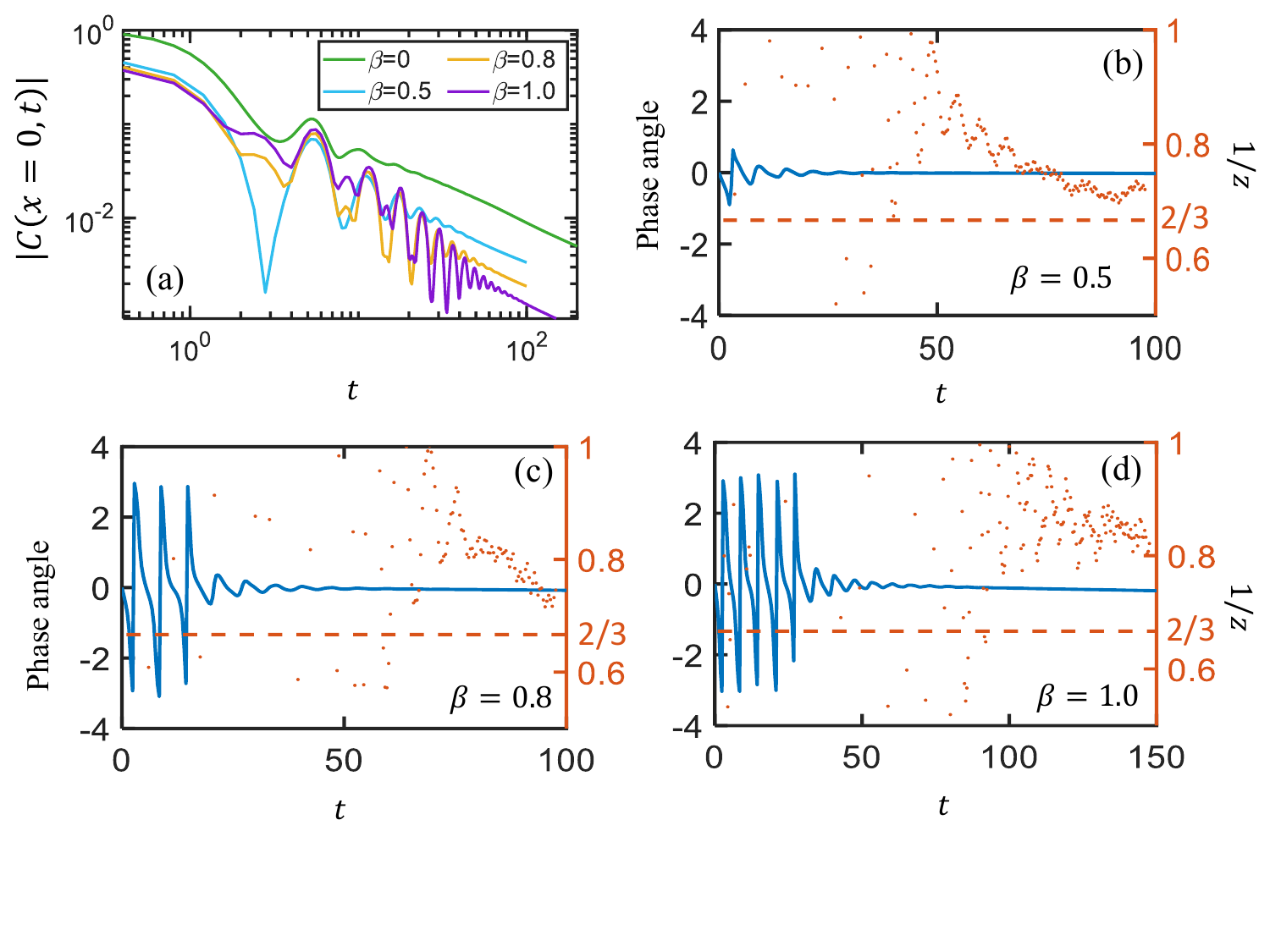}
  \vskip -1.5cm
  \caption{Energy autocorrelator from infinite to moderately high temperatures.  (a) On-site correlator for $\beta=0,0.5,0.8,1.0$, with a maximum $\chi=512,384,384,512$, respectively. One can observe the transition from damped oscillatory behavior at short times to a power-law decay at later times. The crossover timescale increases rapidly with decreasing temperature. (b)--(d) Blue curves show the phase angle of the correlator, i.e. $\mathrm{Arg}[C(x=0,t)]$, while orange points show the running exponent $1/z_\eff(t)$. The onset of hydrodynamic behavior can be observed both as a disappearance of the phase winding and as a ``clustering'' of $1/z$ points. (These two criteria give qualitatively the same trend of the corresponding crossover time with temperature, although quantitatively they differ since the phase winding occurs somewhat earlier than the clustering of $1/z$ points.) The late-time trend of $1/z$ is consistent with it approaching the KPZ value of $2/3$ (dashed red line).}
  \label{fig:highT}
\end{figure*}

The data discussed below are obtained from large-scale tensor network simulations of chains up to $L=400$ using the ITensor library~\cite{itensor}.
The operator $h_x$ is expressed as a matrix product operator (MPO) and undergoes time evolution using time-evolving block decimation (TEBD)~\cite{Vidal07} with fourth-order Trotterization with a time step $\delta t=0.2$. Separately, the thermal density matrix $e^{-\beta H}$ is obtained by evolving an identity MPO with imaginary time $\beta$. The projection is realized as a product of two-site gates $\mathcal{P}=\prod_x (1-n_x n_{x+1})$ applied on the MPOs. 
At high temperatures, the thermal density matrix is close to the identity (within the constrained Hilbert space), and the dynamics are mostly just the growth of the energy operator. At low temperatures, the contribution from the high-energy states is suppressed by the gap, and the dynamics are dominated by the ground-state properties.
As a result, the simulation is mostly challenging at intermediate temperatures, where entanglement grows fastest. This prevents us from reliably simulating $1<\beta<3$. For $\beta\leq1$ and $\beta\geq 3$, we keep a maximum bond dimension $\chi$ between 256 and 512 to ensure convergence, which is listed in the figure captions. The convergence of a representative case $\beta=1$ is shown in the appendix.

We implement several improvements to accelerate and increase the accuracy of the simulations: 
(1) Due to the constraint that no neighboring spins can be simultaneously up, one can group two neighboring spin-half into a spin-one, which imposes the constraint on half of the bonds and reduces the dimension of the local Hilbert space~\cite{Ljubotina2023Superdiff,Lin2019}. 
(2) For the on-site energy correlator $\langle h_0(t) h_0(0)\rangle _\beta$, it is equal to $\langle h_0(t/2) h_0(-t/2)\rangle_\beta$ because of time-translation invariance. Since $h_0(-t/2)$ is the complex conjugate of $h_0(t/2)$, one only needs to evolve half of the time to evaluate Eq.~\eqref{eq:correlator}~\cite{barthel2013precise,karrasch2013reducing}, which also doubles the effective time step $\delta t$ to 0.4. 
(3) Since the thermal density matrix $e^{-\beta H}$ commutes with the time evolution $e^{-iHt}$, $h_x$ can be contracted with $e^{-\beta H}$ before time evolution. We find this is especially useful at low temperatures (large $\beta$), where the entanglement growth with time can be substantially reduced compared with time evolution prior to the thermal tracing. 
(4) At later times when the bond dimension saturates, we take advantage of randomized singular value decomposition (SVD) to speed up the simulation~\cite{Ljubotina2023Superdiff,Tamascelli_2015,Halko_2011}. 
One potential issue with randomized SVD is that it may fail to target some small singular values, especially when the singular value spectrum decays slowly. This issue can be mitigated with power iteration that amplifies the difference in singular values, and with oversampling more singular values than the target number, to ensure those within the target number are accurately captured.
In practice, we find no difference between regular SVD and randomized SVD with one power iteration and oversampling of 10 singular values.

\begin{figure*}[t]
  \centering
  \includegraphics[width=0.8\textwidth]{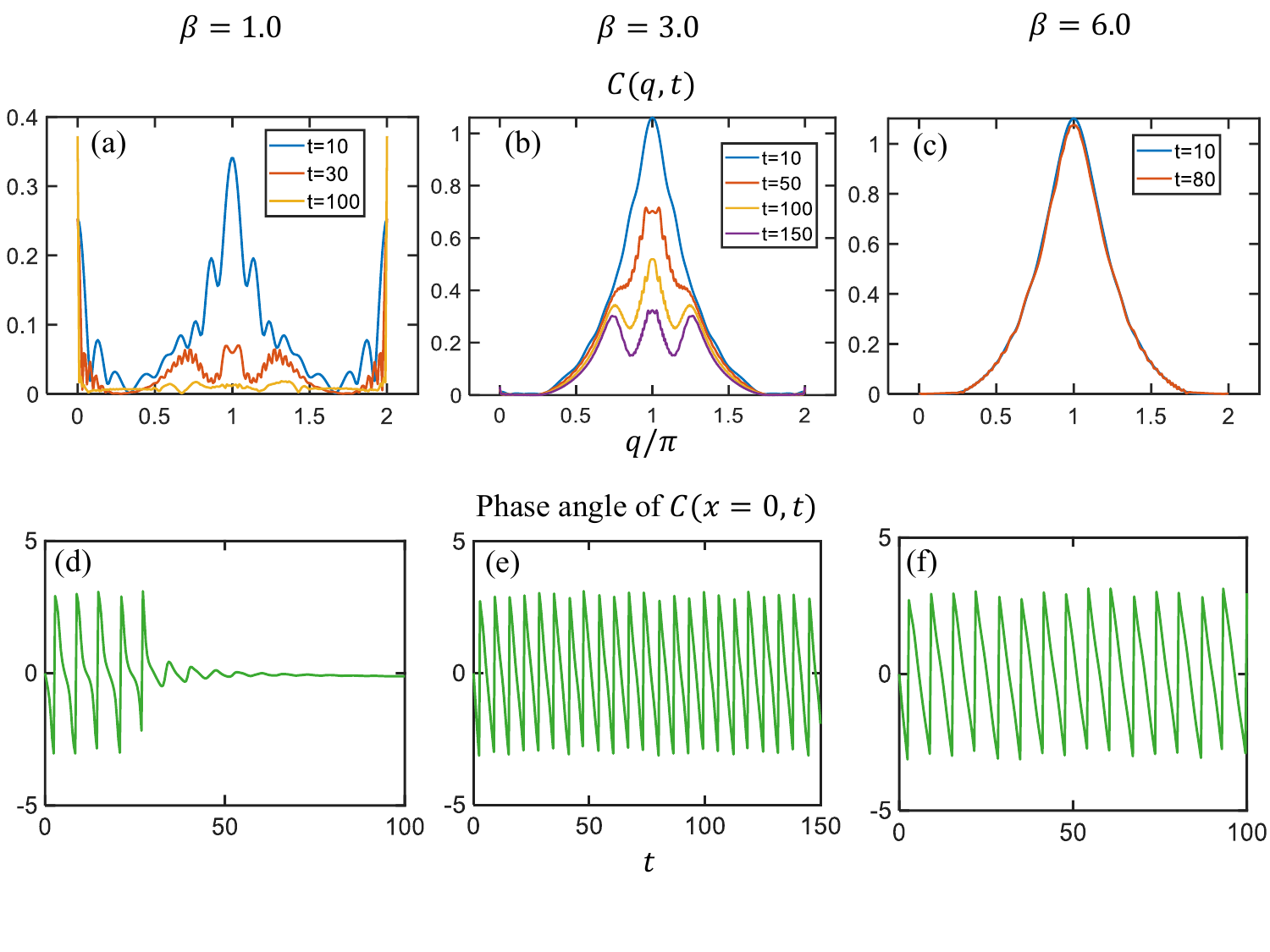}
  \vskip -1.0cm
  \caption{Momentum-space and phase diagnostics of the finite-temperature crossover in the undeformed PXP chain with $\chi=256$. Top row: $C(q,t)$ at representative times. Bottom row: phase angle of the on-site correlator $C(0,t)$. At $\beta=1$ the early-time dynamics is dominated by a peak near $q=\pi$, which gives way at later times to a weight near $q=0$; simultaneously, the phase ceases to wind. At $\beta=3$ and $6$, the dynamics remain dominated by the $q=\pi$ sector over the full simulated time window.}
  \label{fig:cqt}
\end{figure*}

\section{Finite-temperature crossover in energy dynamics}
\label{sec:Results}

In Fig~\ref{fig:highT}(a), we show the on-site correlator $C(0,t)$ for a range of temperatures.
For $\beta=0$, we recover the behavior of Ref.~\onlinecite{Ljubotina2023Superdiff}, with a real correlator showing a power-law decay after a short-time transient with a few damped oscillations.
By contrast, at finite temperature, we observe a parametrically long, coherent regime at short times before hydrodynamics sets in. 
In this coherent regime, the correlator is complex and its phase winds in time, see \cref{fig:highT}(b)--(d). 
As explained in the next section, in this regime the dynamics is dominated by magnons near momentum $\pi$, where the single-magnon dispersion has a minimum.

At later times, the phase winding ceases (the correlator becomes predominantly real), and a power-law decay is recovered. 
This asymptotically real behavior is expected on general grounds. At late times, hydrodynamics predicts a nonoscillatory power-law decay of the symmetrized autocorrelation function, $C_{\mathrm{sym}}(0,t)=\frac{1}{2}\langle\{h_0(t),h_0(0)\}\rangle\sim t^{-1/z}$. The low-frequency fluctuation--dissipation relation then implies that the non-symmetric correlator $C(0,t)$ considered here is such that $\operatorname{Im}C(0,t)\propto\partial_t C_{\mathrm{sym}}(0,t)$, so the imaginary part decays by one additional power of time relative to the real part, causing the phase to approach zero.

The running exponent $1/z_\eff(t)$ starts clustering around an average that slowly drifts down to a value consistent with KPZ.
We observe that the crossover to the hydrodynamic regime occurs for times $\tau$ that increase rapidly as the temperature is lowered.
Due to the finite time window available in numerics, this means that the identification of a hydrodynamic regime and a corresponding exponent $z$ becomes more difficult at lower temperatures.
In fact, for $\beta \gtrsim 3$, $\tau$ becomes larger than the available time window, and we cannot observe the late-time hydrodynamic regime at all.

Nevertheless, we were able to observe the crossover clearly for a range of moderately high temperatures: \Cref{fig:highT}(a) shows the on-site correlator for $\beta=0.5,0.8,1.0$. As the temperature is lowered from infinite temperature into this regime, the coherent short-time regime becomes increasingly pronounced, but the long-time power-law tail still remains accessible. The corresponding phase dynamics and instantaneous exponent are displayed in \cref{fig:highT}(b)--(d). At each temperature, the phase winds through the early-time coherent regime and then settles as the oscillations die out. At a later time, the running exponent $1/z_\eff(t)$ starts clustering around a mean value and drifting downward toward the superdiffusive value $2/3$. Note that the clustering of the running exponent occurs somewhat later than the cessation of phase winding, but both timescales show a similar temperature dependence.

It is also useful to examine the momentum-space correlator shown in \cref{fig:cqt} for $\beta=1,3,6$. At short times, $C(q,t)$ is peaked near $q=\pi$, reflecting the minimum of the magnon band at that momentum (see next section). At later times, once the hydrodynamic regime sets in, the spectral weight shifts toward $q=0$, as expected for the transport of the conserved total energy $\sum_x h_x$, which is a translationally invariant quantity. This transfer is clearly visible at $\beta=1$, where the peak moves from $q=\pi$ to $q=0$ while the phase of the on-site correlator $C(0,t)$ ceases to wind. Upon cooling to $\beta=3$ and $6$, the $q=\pi$ peak remains dominant throughout the accessible time window, and the phase continues to wind, showing that the crossover time has moved beyond the numerically accessible range. 

\begin{figure}[t!]
  \centering
  \includegraphics[width=0.8\columnwidth]{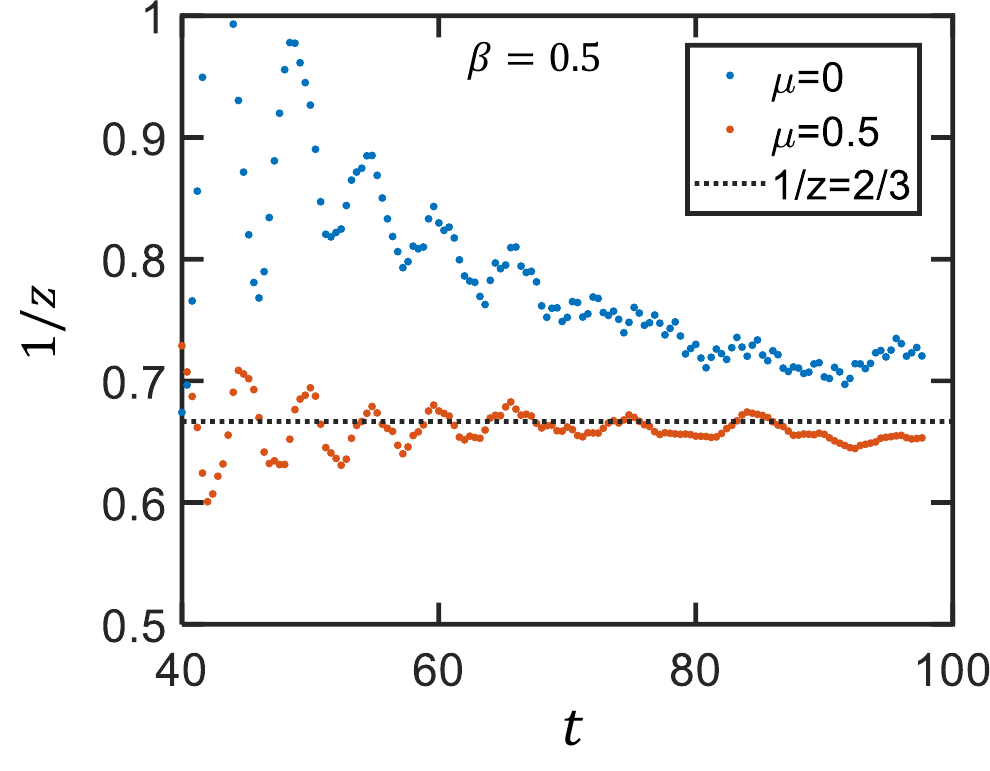}
  \vskip -0.3cm
  \caption{Effect of the PNP deformation on the running exponent at intermediate temperature $\beta=0.5$ with $\chi=384$. The moderate deformation $\mu=0.5$ yields a substantially cleaner plateau near the KPZ value $1/z=2/3$ than the undeformed case $\mu=0$.}
  \label{fig:pnp05}
\end{figure}

Having characterized the undeformed chain, we briefly examine whether the same finite-temperature phenomenology persists under the PNP deformation (Eq.~\ref{PNP}), motivated by the infinite-temperature results of Ref.~\onlinecite{Ljubotina2023Superdiff}, which found that taking positive $\mu$ improves the convergence to the KPZ value of the dynamical exponent.
We have confirmed numerically that this conclusion still applies at finite-$T$: in Fig.~\ref{fig:pnp05}, we show the critical exponent for $\mu=0.5$ and $\beta=0.5$, which stabilizes close to the KPZ value at much shorter times than for the undeformed model.

\section{Magnon description of the coherent regime}
\label{sec:coherent}

We now show how the short-time coherent dynamics can be understood analytically.
As we approach low temperatures, we expect the dynamics to be dominated by low-lying excitations above the ground state.
For PXP, it is well known that these excitations are magnons, with a dispersion minimum at momentum $\pi$~\cite{IadecolaMagnons}.
This picture is closely analogous to the finite-temperature dynamics of $\sigma^z$ in the paramagnetic phase of the transverse-field Ising model (TFIM), as analyzed by Sachdev and Young~\cite{SachdevYoung1997}. In the TFIM $H = -J \sum_i \sigma^z_i \sigma^z_{i+1} - h \sum_i \sigma^x_i $, the single-magnon excitation is odd under the Ising symmetry, so it appears in the two-point function for $\sigma^z$ but not for $\sigma^x$. By contrast, the PXP chain has no Ising symmetry and the single-magnon sector can therefore contribute to the two-point function of any generic local operator, including the energy density considered here.

Guided by Ref.~\onlinecite{SachdevYoung1997}, we write the coherent part of the correlator in the factorized form
\begin{equation}
C^{\mathrm{coh}}(x,t) = K(x,t) R_{\beta}(x,t),
\label{eq:cohfactor}
\end{equation}
where
\begin{equation}
K(x,t) = \int_{-\pi}^{\pi} \frac{{\rm d}p}{2\pi}\, D(p)\, e^{\ii p x - \ii \epsilon_p t}
\label{eq:kernel}
\end{equation}
is the zero-temperature propagation kernel of the single-magnon band with dispersion $\epsilon_p$, and $R_{\beta}(x,t)$ encodes finite-temperature corrections. Here $\epsilon_p$ is the magnon dispersion and $D(p)$ is a form factor that depends on the overlap of the local operator with one-magnon states. In the TFIM, the corresponding kernel can be obtained analytically and expressed in terms of a modified Bessel function~\cite{SachdevYoung1997}, whereas in the PXP chain, there is no known analytical form for it. One important point for us later will be that the magnon band in PXP has its minimum at $p=\pi$, not at $p=0$ as in the TFIM.

\begin{figure}[t!]
  \centering
  \includegraphics[width=\columnwidth]{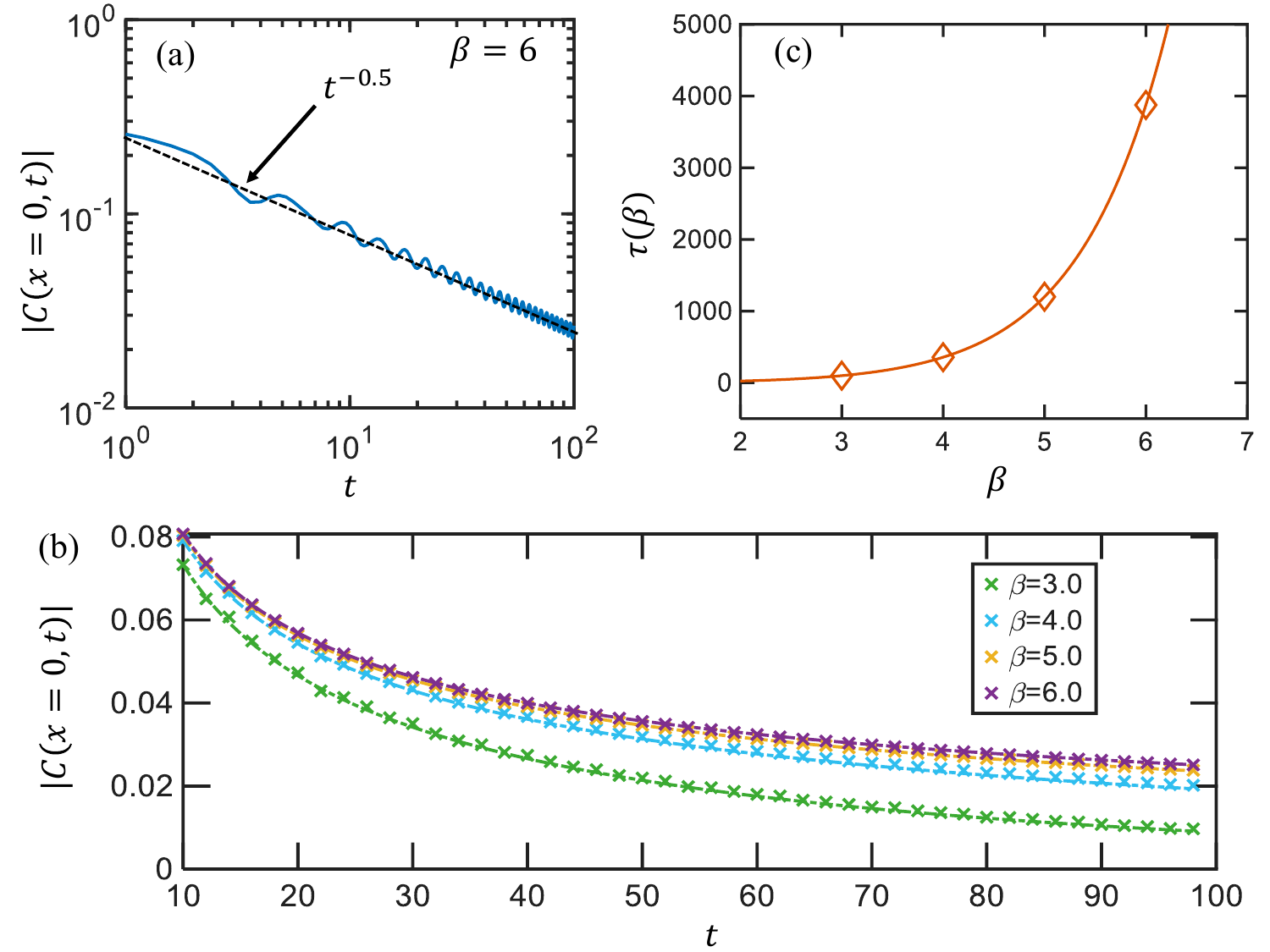}
  \caption{Low-temperature coherent regime in the PXP chain with $\chi=384$. (a) At $\beta=6$, the envelope of the on-site correlator follows the expected $t^{-1/2}$ behavior. (b) For $\beta=3,4,5,6$, the on-site correlators (data points) are well fit by the damped oscillatory form in \cref{eq:fitform} (lines). (c) The activation time $\tau(\beta)=1.7 \beta e^{\Delta \beta}$ as extracted from the fitting in (b). The diamonds correspond to the four temperatures in (b).}
  \label{fig:lowT}
\end{figure}

Expanding the dispersion close to its minimum, we write
\begin{equation}
\epsilon_p  \simeq \Delta + \frac{1}{2}\epsilon'' \delta p^2,
\qquad \delta p \equiv p-\pi,
\label{eq:dispersion}
\end{equation}
with $\Delta = \epsilon_{p=\pi}$ the magnon gap.
At $x=0$, the integral over momentum in Eq.~\ref{eq:kernel} is controlled by the stationary phase at the band minimum, leading to
\begin{equation}
\begin{split}
K(0,t) &\simeq \int \frac{{\rm d}\delta p}{2\pi}\, D(\pi)\,
 e^{-\ii \left(\Delta + \frac{1}{2}\epsilon'' \delta p^2\right)t} \\
&= \frac{D(\pi)\, e^{-\ii \Delta t - \ii \pi/4}}{\sqrt{2\pi \epsilon'' t}} \propto \frac{e^{-\ii \Delta t}}{\sqrt{t}}.
\end{split}
\label{eq:stationaryphase}
\end{equation}
Thus, the coherent short-time signal has the characteristic form $e^{-\ii \Delta t}/\sqrt{t}$, in agreement with the numerics (see \cref{fig:highT,fig:cqt} for the phase winding and \cref{fig:lowT} for the envelope decaying as $t^{-1/2}$). The $t^{-1/2}$ envelope superficially resembles diffusion, but the oscillatory numerator makes clear that the dynamics is coherent.

Going away from zero temperature, the main effect is to suppress this coherent contribution through a relaxation factor. For same-site correlations, based on Sachdev and Young, we expect
\begin{equation}
R_{\beta}(0,t) \simeq e^{-t/\tau(\beta)},
\label{eq:RT}
\end{equation}
whereas in the TFIM it was found that $\tau=(\beta\pi/2)e^{ \Delta\beta}$~\cite{SachdevYoung1997}. By analogy, in the PXP chain, we expect an activated form
\begin{equation}
\tau(\beta) \propto \beta e^{\Delta \beta},
\label{eq:tau}
\end{equation}
with $\Delta$ set by the magnon gap. Here we will use $\Delta=0.97$ as obtained by our density-matrix renormalization group calculations, which is close to the value of 0.9682 reported in Ref.~\cite{IadecolaMagnons}. Combining \cref{eq:stationaryphase,eq:RT} then gives the following prediction for the on-site correlator in the coherent regime:
\begin{equation}
C(x=0,t) \sim A_\beta\, \frac{e^{-\ii \omega t}}{\sqrt{t}}\, e^{-t/\tau(\beta)},
\label{eq:fitform}
\end{equation}
where $\omega=\Delta$ and $A_\beta$ is a nonuniversal amplitude.

\Cref{fig:lowT} shows that the low-temperature data are well described by precisely this picture. Panel (a) displays the magnitude of the on-site correlator at $\beta=6$, whose envelope follows $t^{-1/2}$ over nearly the full accessible range. 
We further fit the magnitude of the on-site correlator across a range of temperatures ($\beta=3,4,5,6$) with the theoretical form in \cref{eq:fitform}. To remove small oscillations in the data that exceed the theory, we apply a Gaussian filter before fitting. We fix the amplitudes $A_\beta$ by matching the data at $t=10$ and then use a single global prefactor in \cref{eq:tau}. A prefactor of 1.7 gives excellent agreement between theory and numerics across all temperatures, as displayed in panel (b). This leads to an activation time $\tau(\beta)=1.7 \beta e^{\Delta \beta}$, plotted in panel (c). 
We note that for $\beta=3,4$, $\tau(\beta)$ is of the same magnitude as our simulation time $t_\text{max}$, and the decaying factor $e^{-t/\tau(\beta)}$ has a visible effect. On the other hand, for $\beta=5,6$, $\tau(\beta)$ is large compared to $t_\text{max}$, and the decay is dominated by $1/\sqrt{t}$.
We also extracted the period of oscillations $T = 2\pi/\omega$ from the phase winding (see Fig~\ref{fig:cqt}.(d-f)), and we obtained $T \approx 6.4$, which is close to the saddle point prediction $T = 2\pi / \Delta \approx 6.48$.

Finally, we note that for negative $\mu$, the PNP deformation of Eq.~\ref{PNP} can close the magnon gap at a quantum critical point separating the paramagnet from an ordered phase~\cite{FendleySachdev}. By analogy with the TFIM, we expect the dynamics near this point to be governed by quantum critical behavior within a critical fan, where the activated form of $\tau$ is replaced by the scaling relation $\tau \propto \beta$. We leave a detailed study of this regime to future work.

\section{Discussion and outlook}
\label{sec:discussion}

The main message of this work is that finite temperature reconciles two dynamical facets of the PXP chain: a short-time coherent regime dominated by $\pi$ magnons and a high-temperature regime with superdiffusive energy transport. The coherent regime is marked by complex oscillations with a $t^{-1/2}$ envelope, and by a pronounced peak in $C(q,t)$ near $q=\pi$, whereas the late-time regime is carried by long wavelengths near $q=0$ and exhibits an effective exponent consistent with KPZ superdiffusion whenever the accessible time window extends beyond the thermalization scale $\tau(\beta)$.

Cooling does not appear to qualitatively alter the eventual transport channel, but it does lengthen the pre-hydrodynamic window controlled by the $\pi$-magnon branch. At a practical level, the rapid, activated growth of $\tau(\beta)$ upon cooling means that the eventual late-time transport regime quickly moves beyond the numerically accessible time window with matrix product states.

An important open question is whether the same magnon sector that controls the short-time signal also plays a direct role in producing the late-time superdiffusive regime, e.g. through magnon-magnon scattering. Our results do not settle that issue, but they strongly constrain any viable mechanism. The hydrodynamic regime does not emerge from featureless local relaxation; rather, it is reached through a temperature-dependent transfer of spectral weight out of a sharply identifiable $q=\pi$ coherent sector. Any microscopic theory of superdiffusion in the PXP chain should therefore account for both the activated crossover scale in \cref{eq:tau} and the observed handoff of spectral weight from $q=\pi$ to $q=0$.

Further, our results highlight an instructive contrast with the transverse-field Ising model. At short times and finite temperature, the PXP energy correlator is dominated by a single magnon band and therefore takes the same functional form as the $\sigma^z$ correlator in the TFIM, namely $e^{-i\Delta t-t/\tau(\beta)}/\sqrt{t}$. The long-time behavior, however, is qualitatively different. In the PXP chain, the correlator crosses over to a power-law tail $t^{-1/z}$ characteristic of superdiffusive energy transport. In the TFIM, by contrast, the $\sigma^z$ correlator remains dominated at late times by the exponential factor $e^{-t/\tau(\beta)}$, which is allowed because $\sigma^z$ has no overlap with the conserved energy density by Ising symmetry. Understanding why similar short-time magnon physics leads to qualitatively different late-time behavior may help isolate the microscopic ingredients required for superdiffusion.

More broadly, our results place the PXP chain alongside a growing class of constrained systems in which sharply identifiable microscopic structures coexist with nontrivial hydrodynamics~\cite{Singh_2021,Pal22}. This makes finite-temperature constrained dynamics a promising arena for developing a microscopic theory of anomalous transport and for connecting it with tensor-network numerics, scar physics, and future experiments on Rydberg platforms.

\begin{acknowledgments}
We thank Maksym Serbyn, Steven White and Thomas Iadecola for insightful discussions. Work by T.S. was supported by the U.S. Department of Energy, Office of Science, Office of Basic Energy Sciences under Early Career Research Program Award Number DE-SC0025568. T.S. gratefully acknowledges the Institute of Science and Technology Austria (ISTA) for hosting him during the initial stages of this project.
S.J. is supported by the U.S. Department of Energy, Office of Science, Basic Energy Sciences, Materials Sciences and Engineering Division, under contract DE-AC02-76SF00515. 
M.L. acknowledges support by the Deutsche Forschungsgemeinschaft (DFG, German Research Foundation) under Germany's Excellence Strategy -- EXC-2111 -- 390814868. J.-Y.D.~acknowledges funding from the European Union's Horizon 2020 research and innovation programme under the Marie Sk\l odowska-Curie Grant Agreement No.~101034413.
\end{acknowledgments}

\appendix*
\section{Convergence of numerical data}
\begin{figure}[h]
  \begin{center}
  \vskip -0.3cm
  \includegraphics[width= 1.0\linewidth]{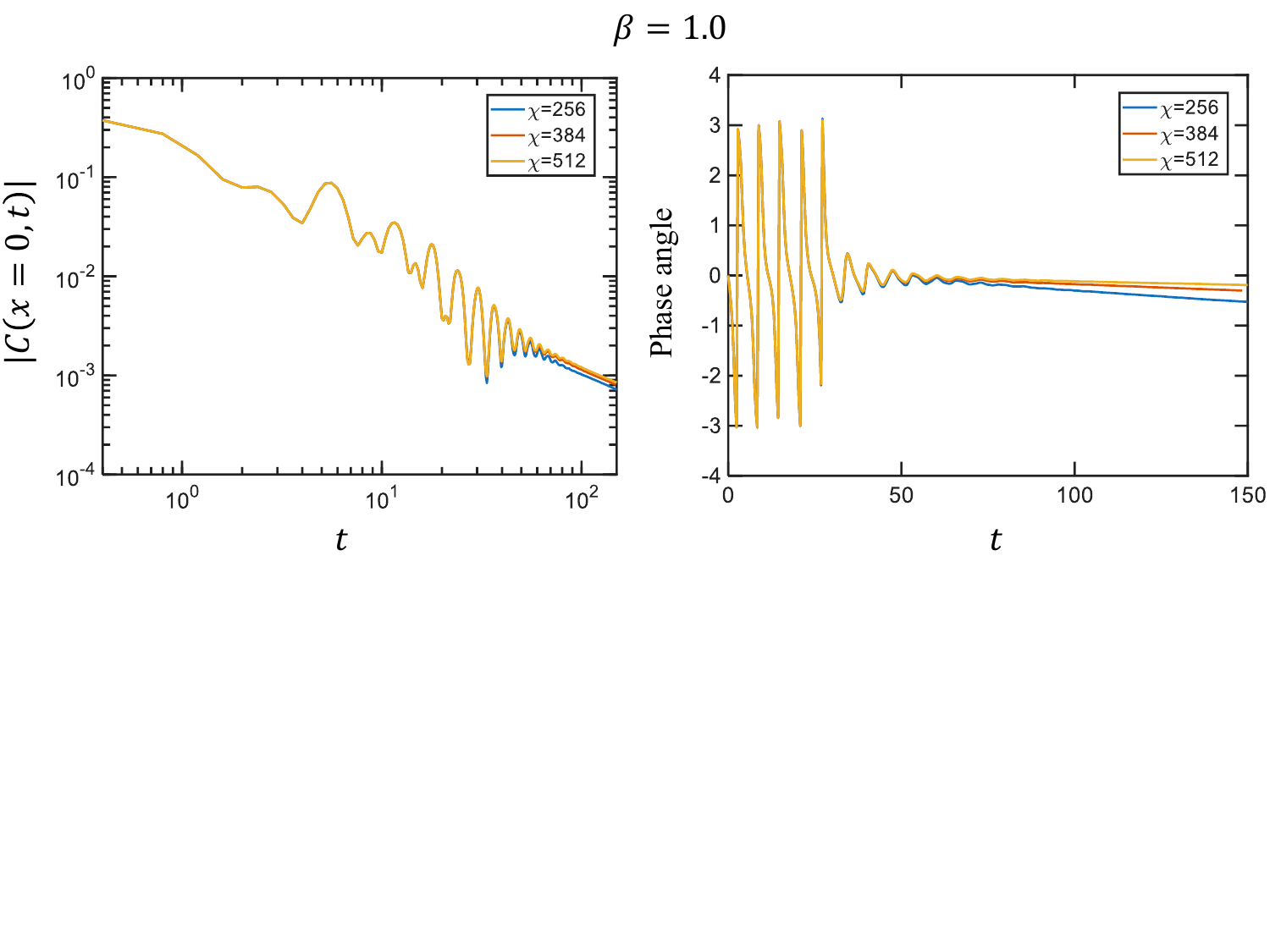}
  \vskip -3cm
  \caption{Onsite energy autocorrelator $|C(x=0,t)|$ (left panel) and its phase angle (right panel) at $\beta=1.0$, at bond dimension $\chi=256,384,512$.}
  \label{fig:convg}
  \end{center}
\end{figure}

For the simulations throughout this work, we compute the results with different bond dimensions to ensure they have converged.
In Fig.~\ref{fig:convg}, we show a representative case  $\beta=1.0$. Increasing the maximum bond dimension $\chi$ only induces minor quantitative changes in the data, which is hardly visible from $\chi=384$ to $\chi=512$. The qualitative features, including the crossover time and the long-time superdiffusive transport, have already converged even at the smallest $\chi$.

\bibliography{refs}

@article{Ljubotina2023Superdiff,
  title = {Superdiffusive Energy Transport in Kinetically Constrained Models},
  author = {Ljubotina, Marko and Desaules, Jean-Yves and Serbyn, Maksym and Papi\'c, Zlatko},
  journal = {Phys. Rev. X},
  volume = {13},
  issue = {1},
  pages = {011033},
  year = {2023},
  doi = {10.1103/PhysRevX.13.011033}
}

@misc{ThuillierScaffidi2026AC,
  author        = {Thuillier, Davis and Scaffidi, Thomas},
  title         = {{{AC} Fingerprints of {2D} Electron Hydrodynamics: Superdiffusion and Drude Weight Suppression}},
  year          = {2026},
  eprint        = {2603.15737},
  archivePrefix = {arXiv},
  primaryClass  = {cond-mat.str-el},
  doi           = {10.48550/arXiv.2603.15737}
}

@article{Hui_2025,
doi = {10.1088/1361-648X/adfbcd},
url = {https://doi.org/10.1088/1361-648X/adfbcd},
year = {2025},
month = {sep},
publisher = {IOP Publishing},
volume = {37},
number = {36},
pages = {363001},
author = {Hui, Aaron and Skinner, Brian},
title = {Hydrodynamics of the electronic Fermi liquid: a pedagogical overview},
journal = {Journal of Physics: Condensed Matter}
}

@article{Lucas_2018,
doi = {10.1088/1361-648X/aaa274},
url = {https://doi.org/10.1088/1361-648X/aaa274},
year = {2018},
month = {jan},
publisher = {IOP Publishing},
volume = {30},
number = {5},
pages = {053001},
author = {Lucas, Andrew and Fong, Kin Chung},
title = {Hydrodynamics of electrons in graphene},
journal = {Journal of Physics: Condensed Matter}
}

@article{FritzScaffidi,
   author = "Fritz, L. and Scaffidi, T.",
   title = "Hydrodynamic Electronic Transport", 
   journal= "Annual Review of Condensed Matter Physics",
   year = "2024",
   volume = "15",
   number = "Volume 15, 2024",
   pages = "17-44",
   doi = "https://doi.org/10.1146/annurev-conmatphys-040521-042014",
   url = "https://www.annualreviews.org/content/journals/10.1146/annurev-conmatphys-040521-042014",
   publisher = "Annual Reviews",
   issn = "1947-5462",
   type = "Journal Article",
  }

@article{Bertini2021RMP,
  title = {Finite-temperature transport in one-dimensional quantum lattice models},
  author = {Bertini, Bruno and Heidrich-Meisner, Fabian and Karrasch, Christoph and Prosen, Toma\v{z} and Steinigeweg, Robin and \v{Z}nidari\v{c}, Marko},
  journal = {Rev. Mod. Phys.},
  volume = {93},
  issue = {2},
  pages = {025003},
  year = {2021},
  doi = {10.1103/RevModPhys.93.025003}
}

@article{Weiner2020KPZ,
  title = {High-temperature spin dynamics in the {Heisenberg} chain: Magnon propagation and emerging {Kardar-Parisi-Zhang} scaling in the zero-magnetization limit},
  author = {Weiner, Felix and Schmitteckert, Peter and Bera, Soumya and Evers, Ferdinand},
  journal = {Phys. Rev. B},
  volume = {101},
  issue = {4},
  pages = {045115},
  year = {2020},
  doi = {10.1103/PhysRevB.101.045115}
}

@article{SachdevYoung1997,
  title = {Low Temperature Relaxational Dynamics of the {Ising} Chain in a Transverse Field},
  author = {Sachdev, Subir and Young, A. P.},
  journal = {Phys. Rev. Lett.},
  volume = {78},
  issue = {11},
  pages = {2220--2223},
  year = {1997},
  doi = {10.1103/PhysRevLett.78.2220}
}

@article{Bernien2017,
  title = {Probing many-body dynamics on a 51-atom quantum simulator},
  author = {Bernien, Hannes and Schwartz, Sylvain and Keesling, Alexander and Levine, Harry and Omran, Ahmed and Pichler, Hannes and Choi, Soonwon and Zibrov, Alexander S. and Endres, Manuel and Greiner, Markus and Vuleti\'c, Vladan and Lukin, Mikhail D.},
  journal = {Nature},
  volume = {551},
  pages = {579--584},
  year = {2017},
  doi = {10.1038/nature24622}
}

@article{Bluvstein2021,
author = {D. Bluvstein  and A. Omran  and H. Levine  and A. Keesling  and G. Semeghini  and S. Ebadi  and T. T. Wang  and A. A. Michailidis  and N. Maskara  and W. W. Ho  and S. Choi  and M. Serbyn  and M. Greiner  and V. Vuleti\'{c}  and M. D. Lukin},
title = {Controlling quantum many-body dynamics in driven {Rydberg} atom arrays},
journal = {Science},
volume = {371},
number = {6536},
pages = {1355-1359},
year = {2021},
doi = {10.1126/science.abg2530},
URL = {https://www.science.org/doi/abs/10.1126/science.abg2530},
eprint = {}
}

@article{Bulchandani_2020,
  title = {{Kardar}-{Parisi}-{Zhang} universality from soft gauge modes},
  author = {Bulchandani, Vir B.},
  journal = {Phys. Rev. B},
  volume = {101},
  issue = {4},
  pages = {041411(R)},
  numpages = {6},
  year = {2020},
  month = {Jan},
  publisher = {American Physical Society},
  doi = {10.1103/PhysRevB.101.041411},
  url = {https://link.aps.org/doi/10.1103/PhysRevB.101.041411}
}

@article{Bulchandani_2020_2,
	doi = {10.1073/pnas.1916213117},
  
	url = {https://doi.org/10.1073%2Fpnas.1916213117},
  
	year = 2020,
	month = {may},
  
	publisher = {Proceedings of the National Academy of Sciences},
  
	volume = {117},
  
	number = {23},
  
	pages = {12713--12718},
  
	author = {Vir B. Bulchandani and Christoph Karrasch and Joel E. Moore},
  
	title = {Superdiffusive transport of energy in one-dimensional metals},
  
	journal = {Proceedings of the National Academy of Sciences}
}

@article{Bulchandani_2021,
	doi = {10.1088/1742-5468/ac12c7},
  
	url = {https://doi.org/10.1088%2F1742-5468%2Fac12c7},
  
	year = 2021,
	month = {aug},
  
	publisher = {{IOP} Publishing},
  
	volume = {2021},
  
	number = {8},
  
	pages = {084001},
  
	author = {Vir B Bulchandani and Sarang Gopalakrishnan and Enej Ilievski},
  
	title = {Superdiffusion in spin chains},
  
	journal = {Journal of Statistical Mechanics: Theory and Experiment}
}

@article{ChandranReview,
   author = "Chandran, Anushya and Iadecola, Thomas and Khemani, Vedika and Moessner, Roderich",
   title = "Quantum Many-Body Scars: A Quasiparticle Perspective", 
   journal= "Annual Review of Condensed Matter Physics",
   year = "2023",
   volume = "14",
   number = "Volume 14, 2023",
   pages = "443-469",
   doi = "https://doi.org/10.1146/annurev-conmatphys-031620-101617",
   publisher = "Annual Reviews",
   issn = "1947-5462",
   type = "Journal Article",
  }

@article{De_Nardis_2019,
  title = {Anomalous Spin Diffusion in One-Dimensional Antiferromagnets},
  author = {De Nardis, Jacopo and Medenjak, Marko and Karrasch, Christoph and Ilievski, Enej},
  journal = {Phys. Rev. Lett.},
  volume = {123},
  issue = {18},
  pages = {186601},
  numpages = {6},
  year = {2019},
  month = {Oct},
  publisher = {American Physical Society},
  doi = {10.1103/PhysRevLett.123.186601},
  url = {https://link.aps.org/doi/10.1103/PhysRevLett.123.186601}
}

@article{De_Nardis_2020,
  title = {Universality Classes of Spin Transport in One-Dimensional Isotropic Magnets: The Onset of Logarithmic Anomalies},
  author = {De Nardis, Jacopo and Medenjak, Marko and Karrasch, Christoph and Ilievski, Enej},
  journal = {Phys. Rev. Lett.},
  volume = {124},
  issue = {21},
  pages = {210605},
  numpages = {6},
  year = {2020},
  month = {May},
  publisher = {American Physical Society},
  doi = {10.1103/PhysRevLett.124.210605},
  url = {https://link.aps.org/doi/10.1103/PhysRevLett.124.210605}
}

@article{FendleySachdev,
	Author = {Fendley, Paul and Sengupta, K. and Sachdev, Subir},
	Doi = {10.1103/PhysRevB.69.075106},
	Issue = {7},
	Journal = {Phys. Rev. B},
	Month = {Feb},
	Numpages = {15},
	Pages = {075106},
	Publisher = {American Physical Society},
	Title = {Competing density-wave orders in a one-dimensional hard-boson model},
	Url = {https://link.aps.org/doi/10.1103/PhysRevB.69.075106},
	Volume = {69},
	Year = {2004}}

@article{Friedman_2020,
  title = {Diffusive hydrodynamics from integrability breaking},
  author = {Friedman, Aaron J. and Gopalakrishnan, Sarang and Vasseur, Romain},
  journal = {Phys. Rev. B},
  volume = {101},
  issue = {18},
  pages = {180302(R)},
  numpages = {6},
  year = {2020},
  month = {May},
  publisher = {American Physical Society},
  doi = {10.1103/PhysRevB.101.180302},
  url = {https://link.aps.org/doi/10.1103/PhysRevB.101.180302}
}

@article{Halko_2011,
author = {Halko, N. and Martinsson, P. G. and Tropp, J. A.},
title = {Finding Structure with Randomness: Probabilistic Algorithms for Constructing Approximate Matrix Decompositions},
journal = {SIAM Review},
volume = {53},
number = {2},
pages = {217-288},
year = {2011},
doi = {10.1137/090771806},

URL = { 
        https://doi.org/10.1137/090771806
    
},
eprint = {}
}

@article{IadecolaMagnons,
  title = {Quantum many-body scars from magnon condensation},
  author = {Iadecola, Thomas and Schecter, Michael and Xu, Shenglong},
  journal = {Phys. Rev. B},
  volume = {100},
  issue = {18},
  pages = {184312},
  numpages = {12},
  year = {2019},
  month = {Nov},
  publisher = {American Physical Society},
  doi = {10.1103/PhysRevB.100.184312},
  url = {https://link.aps.org/doi/10.1103/PhysRevB.100.184312}
}

@article{Ilievski_2018,
  title = {Superdiffusion in One-Dimensional Quantum Lattice Models},
  author = {Ilievski, Enej and De Nardis, Jacopo and Medenjak, Marko and Prosen, Toma\ifmmode\check{z}\else\v{z}\fi{}},
  journal = {Phys. Rev. Lett.},
  volume = {121},
  issue = {23},
  pages = {230602},
  numpages = {6},
  year = {2018},
  month = {Dec},
  publisher = {American Physical Society},
  doi = {10.1103/PhysRevLett.121.230602},
  url = {https://link.aps.org/doi/10.1103/PhysRevLett.121.230602}
}

@article{Karrasch_2014,
  title = {Real-time and real-space spin and energy dynamics in one-dimensional spin-$\frac{1}{2}$ systems induced by local quantum quenches at finite temperatures},
  author = {Karrasch, C. and Moore, J. E. and Heidrich-Meisner, F.},
  journal = {Phys. Rev. B},
  volume = {89},
  issue = {7},
  pages = {075139},
  numpages = {12},
  year = {2014},
  month = {Feb},
  publisher = {American Physical Society},
  doi = {10.1103/PhysRevB.89.075139},
  url = {https://link.aps.org/doi/10.1103/PhysRevB.89.075139}
}

@article{Labuhn2016,
	Author = {Labuhn, Henning and Barredo, Daniel and Ravets, Sylvain and de L{\'e}s{\'e}leuc, Sylvain and Macr{\`\i}, Tommaso and Lahaye, Thierry and Browaeys, Antoine},
	Day = {01},
	Journal = {Nature},
	Month = {Jun},
	Pages = {667},
	Publisher = {Nature Publishing Group, a division of Macmillan Publishers Limited. All Rights Reserved. SN -},
	Title = {Tunable two-dimensional arrays of single {Rydberg} atoms for realizing quantum {Ising} models},
	Url = {http://dx.doi.org/10.1038/nature18274},
	Volume = {534},
	Year = {2016}}

@article{Lesanovsky2012,
	Author = {Lesanovsky, Igor and Katsura, Hosho},
	Doi = {10.1103/PhysRevA.86.041601},
	Issue = {4},
	Journal = {Phys. Rev. A},
	Month = {Oct},
	Numpages = {5},
	Pages = {041601(R)},
	Publisher = {American Physical Society},
	Title = {Interacting {Fibonacci} anyons in a {Rydberg} gas},
	Url = {https://link.aps.org/doi/10.1103/PhysRevA.86.041601},
	Volume = {86},
	Year = {2012}}

@article{Lin2019,
  title = {Exact Quantum Many-Body Scar States in the {Rydberg}-Blockaded Atom Chain},
  author = {Lin, Cheng-Ju and Motrunich, Olexei I.},
  journal = {Phys. Rev. Lett.},
  volume = {122},
  issue = {17},
  pages = {173401},
  numpages = {5},
  year = {2019},
  month = {Apr},
  publisher = {American Physical Society},
  doi = {10.1103/PhysRevLett.122.173401},
  url = {https://link.aps.org/doi/10.1103/PhysRevLett.122.173401}
}

@article{Ljubotina_2017,
	doi = {10.1038/ncomms16117},
  
	url = {https://doi.org/10.1038%2Fncomms16117},
  
	year = 2017,
	month = {jul},
  
	publisher = {Springer Science and Business Media {LLC}
},
  
	volume = {8},
	pages=16117,
  
	number = {1},
  
	author = {Marko Ljubotina and Marko {\v{Z}}nidari{\v{c}} and Toma{\v{z}} Prosen},
  
	title = {Spin diffusion from an inhomogeneous quench in an integrable system},
  
	journal = {Nature Communications}
}

@article{Ljubotina_2019,
  title = {{Kardar}-{Parisi}-{Zhang} Physics in the Quantum {Heisenberg} Magnet},
  author = {Ljubotina, Marko and \ifmmode\check{Z}\else\v{Z}\fi{}nidari\ifmmode\check{c}\else\v{c}\fi{}, Marko and Prosen, Toma\ifmmode\check{z}\else\v{z}\fi{}},
  journal = {Phys. Rev. Lett.},
  volume = {122},
  issue = {21},
  pages = {210602},
  numpages = {6},
  year = {2019},
  month = {May},
  publisher = {American Physical Society},
  doi = {10.1103/PhysRevLett.122.210602},
  url = {https://link.aps.org/doi/10.1103/PhysRevLett.122.210602}
}

@article{Lux2014,
  title = {Hydrodynamic long-time tails after a quantum quench},
  author = {Lux, Jonathan and M\"uller, Jan and Mitra, Aditi and Rosch, Achim},
  journal = {Phys. Rev. A},
  volume = {89},
  issue = {5},
  pages = {053608},
  numpages = {8},
  year = {2014},
  month = {May},
  publisher = {American Physical Society},
  doi = {10.1103/PhysRevA.89.053608},
  url = {https://link.aps.org/doi/10.1103/PhysRevA.89.053608}
}

@article{MoudgalyaReview,
	doi = {10.1088/1361-6633/ac73a0},
	url = {https://doi.org/10.1088/1361-6633/ac73a0},
	year = 2022,
	month = {jul},
	publisher = {{IOP} Publishing},
	volume = {85},
	number = {8},
	pages = {086501},
	author = {Sanjay Moudgalya and B Andrei Bernevig and Nicolas Regnault},
	title = {Quantum many-body scars and {Hilbert} space fragmentation: a review of exact results},
	journal = {Reports on Progress in Physics}
}

@article{Pal22,
  title = {Anomalous hydrodynamics in a class of scarred frustration-free Hamiltonians},
  author = {Richter, Jonas and Pal, Arijeet},
  journal = {Phys. Rev. Research},
  volume = {4},
  issue = {1},
  pages = {L012003},
  numpages = {8},
  year = {2022},
  month = {Jan},
  publisher = {American Physical Society},
  doi = {10.1103/PhysRevResearch.4.L012003},
  url = {https://link.aps.org/doi/10.1103/PhysRevResearch.4.L012003}
}

@article{Prosen_2012,
  title = {Diffusive high-temperature transport in the one-dimensional {Hubbard} model},
  author = {Prosen, Toma\ifmmode\check{z}\else\v{z}\fi{} and \ifmmode\check{Z}\else\v{Z}\fi{}nidari\ifmmode\check{c}\else\v{c}\fi{}, Marko},
  journal = {Phys. Rev. B},
  volume = {86},
  issue = {12},
  pages = {125118},
  numpages = {6},
  year = {2012},
  month = {Sep},
  publisher = {American Physical Society},
  doi = {10.1103/PhysRevB.86.125118},
  url = {https://link.aps.org/doi/10.1103/PhysRevB.86.125118}
}

@Article{Scheie2021,
author={Scheie, A.
and Sherman, N. E.
and Dupont, M.
and Nagler, S. E.
and Stone, M. B.
and Granroth, G. E.
and Moore, J. E.
and Tennant, D. A.},
title={Detection of {Kardar}--{Parisi}--{Zhang} hydrodynamics in a quantum {Heisenberg} spin-1/2 chain},
journal={Nature Physics},
year={2021},
month={Jun},
day={01},
volume={17},
number={6},
pages={726-730},
issn={1745-2481},
doi={10.1038/s41567-021-01191-6},
url={https://doi.org/10.1038/s41567-021-01191-6}
}

@article{Serbyn2021,
	Author = {Serbyn, Maksym and Abanin, Dmitry A. and Papi{\'c}, Zlatko},
	Da = {2021/06/01},
	Doi = {10.1038/s41567-021-01230-2},
	Id = {Serbyn2021},
	Isbn = {1745-2481},
	Journal = {Nature Physics},
	Number = {6},
	Pages = {675--685},
	Title = {Quantum many-body scars and weak breaking of ergodicity},
	Ty = {JOUR},
	Url = {https://doi.org/10.1038/s41567-021-01230-2},
	Volume = {17},
	Year = {2021}}

@article{Singh_2021,
  title = {Subdiffusion and Many-Body Quantum Chaos with Kinetic Constraints},
  author = {Singh, Hansveer and Ware, Brayden A. and Vasseur, Romain and Friedman, Aaron J.},
  journal = {Phys. Rev. Lett.},
  volume = {127},
  issue = {23},
  pages = {230602},
  numpages = {6},
  year = {2021},
  month = {Dec},
  publisher = {American Physical Society},
  doi = {10.1103/PhysRevLett.127.230602},
  url = {https://link.aps.org/doi/10.1103/PhysRevLett.127.230602}
}

@article{Spohn_2014,
	doi = {10.1007/s10955-014-0933-y},
  
	url = {https://doi.org/10.1007%2Fs10955-014-0933-y},
  
	year = 2014,
	month = {feb},
  
	publisher = {Springer Science and Business Media {LLC}
},
  
	volume = {154},
  
	number = {5},
  
	pages = {1191--1227},
  
	author = {Herbert Spohn},
  
	title = {Nonlinear Fluctuating Hydrodynamics for Anharmonic Chains},
  
	journal = {Journal of Statistical Physics}
}

@article{Surace2020,
  title = {Lattice Gauge Theories and String Dynamics in {Rydberg} Atom Quantum Simulators},
  author = {Surace, Federica M. and Mazza, Paolo P. and Giudici, Giuliano and Lerose, Alessio and Gambassi, Andrea and Dalmonte, Marcello},
  journal = {Phys. Rev. X},
  volume = {10},
  issue = {2},
  pages = {021041},
  numpages = {14},
  year = {2020},
  month = {May},
  publisher = {American Physical Society},
  doi = {10.1103/PhysRevX.10.021041},
  url = {https://link.aps.org/doi/10.1103/PhysRevX.10.021041}
}

@article{Tamascelli_2015,
  title = {Improved scaling of time-evolving block-decimation algorithm through reduced-rank randomized singular value decomposition},
  author = {Tamascelli, D. and Rosenbach, R. and Plenio, M. B.},
  journal = {Phys. Rev. E},
  volume = {91},
  issue = {6},
  pages = {063306},
  numpages = {12},
  year = {2015},
  month = {Jun},
  publisher = {American Physical Society},
  doi = {10.1103/PhysRevE.91.063306},
  url = {https://link.aps.org/doi/10.1103/PhysRevE.91.063306}
}

@article{Vidal07,
	Author = {Vidal, G.},
	Doi = {10.1103/PhysRevLett.98.070201},
	Issue = {7},
	Journal = {Phys. Rev. Lett.},
	Month = {Feb},
	Numpages = {4},
	Pages = {070201},
	Publisher = {American Physical Society},
	Title = {Classical Simulation of Infinite-Size Quantum Lattice Systems in One Spatial Dimension},
	Volume = {98},
	Year = {2007}}

@article{Znidaric_2011,
  title = {Spin Transport in a One-Dimensional Anisotropic {Heisenberg} Model},
  author = {\ifmmode\check{Z}\else\v{Z}\fi{}nidari\ifmmode\check{c}\else\v{c}\fi{}, Marko},
  journal = {Phys. Rev. Lett.},
  volume = {106},
  issue = {22},
  pages = {220601},
  numpages = {4},
  year = {2011},
  month = {May},
  publisher = {American Physical Society},
  doi = {10.1103/PhysRevLett.106.220601},
  url = {https://link.aps.org/doi/10.1103/PhysRevLett.106.220601}
}

@misc{itensor,
	title={{The ITensor Software Library for Tensor Network Calculations}},
	author={Matthew Fishman and Steven R. White and E. Miles Stoudenmire},
	journal={SciPost Phys. Codebases},
	pages={4},
	year={2022},
	publisher={SciPost},
	doi={10.21468/SciPostPhysCodeb.4},
	url={https://scipost.org/10.21468/SciPostPhysCodeb.4},
}

@article{Dupont2021Crossover,
  title = {Spatiotemporal Crossover between Low- and High-Temperature Dynamical Regimes in the Quantum {Heisenberg} Magnet},
  author = {Dupont, Maxime and Sherman, Nicholas E. and Moore, Joel E.},
  journal = {Phys. Rev. Lett.},
  volume = {127},
  issue = {10},
  pages = {107201},
  year = {2021},
  doi = {10.1103/PhysRevLett.127.107201}
}

@article{DeNardisSolitons2020,
  title = {Superdiffusion from Emergent Classical Solitons in Quantum Spin Chains},
  author = {De Nardis, Jacopo and Gopalakrishnan, Sarang and Ilievski, Enej and Vasseur, Romain},
  journal = {Phys. Rev. Lett.},
  volume = {125},
  issue = {7},
  pages = {070601},
  year = {2020},
  doi = {10.1103/PhysRevLett.125.070601}
}

@article{Wei2022KPZ,
  title = {Quantum gas microscopy of {Kardar-Parisi-Zhang} superdiffusion},
  author = {Wei, David and Rubio-Abadal, Antonio and Ye, Bingtian and Machado, Francisco and Kemp, Jack and Srakaew, Kritsana and Hollerith, Simon and Rui, Jun and Gopalakrishnan, Sarang and Yao, Norman Y. and Bloch, Immanuel and Zeiher, Johannes},
  journal = {Science},
  volume = {376},
  number = {6594},
  pages = {716--720},
  year = {2022},
  doi = {10.1126/science.abk2397}
}

@article{Rosenberg2024Science,
  title = {Dynamics of magnetization at infinite temperature in a {Heisenberg} spin chain},
  author = {E. Rosenberg  and T. I. Andersen  and R. Samajdar  and A. Petukhov  and J. C. Hoke  and D. Abanin  and A. Bengtsson  and I. K. Drozdov  and C. Erickson  and P. V. Klimov  and others},
  journal = {Science},
  volume = {384},
  number = {6691},
  pages = {48--53},
  year = {2024},
  doi = {10.1126/science.adi7877}
}

@article{barthel2013precise,
  title={Precise evaluation of thermal response functions by optimized density matrix renormalization group schemes},
  author={Barthel, Thomas},
  journal={New Journal of Physics},
  volume={15},
  number={7},
  pages={073010},
  year={2013},
  publisher={IOP Publishing},
  doi={10.1088/1367-2630/15/7/073010}
}

@article{karrasch2013reducing,
  title={Reducing the numerical effort of finite-temperature density matrix renormalization group calculations},
  author={Karrasch, C and Bardarson, JH and Moore, JE},
  journal={New Journal of Physics},
  volume={15},
  number={8},
  pages={083031},
  year={2013},
  publisher={IOP Publishing},
  doi={10.1088/1367-2630/15/8/083031}
}

@article{bhakuni2025LGT,
      title={Anomalously fast transport in non-integrable lattice gauge theories}, 
      author={Devendra Singh Bhakuni and Roberto Verdel and Jean-Yves Desaules and Maksym Serbyn and Marko Ljubotina and Marcello Dalmonte},
      year={2025},
      eprint={2509.08889},
      archivePrefix={arXiv},
      primaryClass={cond-mat.quant-gas},
      journal={arXiv e-Prints}
}

@article{park2025graph,
  title = {Graph-theoretical proof of nonintegrability in quantum many-body systems: Application to the PXP model},
  author = {Park, HaRu K. and Lee, SungBin},
  journal = {Phys. Rev. B},
  volume = {111},
  issue = {8},
  pages = {L081101},
  numpages = {6},
  year = {2025},
  month = {Feb},
  publisher = {American Physical Society},
  doi = {10.1103/PhysRevB.111.L081101},
  url = {https://link.aps.org/doi/10.1103/PhysRevB.111.L081101}
}

@book{Sutherland2004,
  title = {Beautiful Models: 70 Years of Exactly Solved Quantum Many-Body Problems},
  author = {Sutherland, Bill},
  year = {2004},
  publisher = {World Scientific},
  address = {Singapore},
  doi = {10.1142/5552}
}

@article{steve-xxz,
  title = {Exact Edge Singularities and Dynamical Correlations in Spin-$1/2$ Chains},
  author = {Pereira, Rodrigo G. and White, Steven R. and Affleck, Ian},
  journal = {Phys. Rev. Lett.},
  volume = {100},
  issue = {2},
  pages = {027206},
  numpages = {4},
  year = {2008},
  month = {Jan},
  publisher = {American Physical Society},
  doi = {10.1103/PhysRevLett.100.027206},
}

\end{document}